\documentclass{article}

\usepackage{ifthen}
\newboolean{neurips_version}
\setboolean{neurips_version}{true}
\ifthenelse{\boolean{neurips_version}}
{\PassOptionsToPackage{numbers, compress}{natbib}
 \usepackage[preprint]{neurips_2022}
}
{
\usepackage{PRIMEarxiv}
 \usepackage[numbers]{natbib}
}

\usepackage[utf8]{inputenc} 
\usepackage[T1]{fontenc}    
\usepackage{hyperref}       
\usepackage{url}            
\usepackage{booktabs}       
\usepackage{amsfonts,amsmath,amsthm,amssymb}       
\usepackage{nicefrac}       
\usepackage{microtype}      
\usepackage{xcolor}         
\usepackage[disable]{todonotes}
\usepackage{enumitem}
\usepackage{caption}

\usepackage{mathtools} 
\usepackage{tikz}
\usetikzlibrary{calc}
\usepackage{pgfplots}
\usepackage{graphicx}

\usepackage{amsmath}
\usepackage{algorithm, algpseudocode}
\usepackage{amsfonts}  

\usepackage[english]{babel}
\usepackage{amsthm}

\usepackage{subfig}
\usepackage[export]{adjustbox}

\usepackage{hhline}
\usepackage{makecell, caption, booktabs}
\usepackage{siunitx}

\usepackage{multirow}

\usepackage{amsmath, nccmath}
\usepackage{bigstrut}

\usepackage{booktabs}

\usepackage{lipsum}
\graphicspath{{media/}}     

\usepackage{CJKutf8}
\usepackage[framed,numbered,autolinebreaks,useliterate]{mcode}

\title{ Privacy-Preserving CNN Training with Transfer Learning: Two Hidden Layers  }

\author{ \href{https://orcid.org/0000-0003-0378-0607}{\includegraphics[scale=0.06]{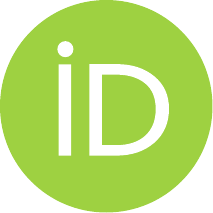}\hspace{1mm}John Chiang} \\                             
                                      \\
	\texttt{john.chiang.smith@gmail.com} 
}

\date{}

\theoremstyle{remark}

\renewcommand{\epsilon}{\varepsilon}

\makeatletter
\def\namedlabel#1#2{\begingroup
    #2%
    \def\@currentlabel{#2}%
    \phantomsection\label{#1}\endgroup
}
\makeatother

\algnewcommand{\LeftComment}[1]{\Statex \(\triangleright\) #1}
\algnewcommand{\LineCommentStep}[1]{\Statex \textbf{[Step #1]:} }
\makeatletter
\newlength{\trianglerightwidth}
\algnewcommand{\LineComment}[1]{\Statex \hskip\ALG@thistlm $\triangleright$ #1}
\algnewcommand{\LineCommentCont}[1]{\Statex \hskip\ALG@thistlm%
  \parbox[t]{\dimexpr\linewidth-\ALG@thistlm}{\hangindent=\trianglerightwidth \hangafter=1 \strut$\triangleright$ #1\strut}}
\algnewcommand{\LeftLineCommentCont}[1]{\Statex \hskip\ALG@thistlm%
  \parbox[t]{\dimexpr\linewidth-\ALG@thistlm}{\leftskip=\algorithmicindent \hangindent=\trianglerightwidth \hangafter=1 \strut$\triangleright$ #1\strut}}

\newcommand{\mysplit}[1]{%
  \begin{tabular}{@{}c@{}}
    #1
  \end{tabular}
  }
  
\begin{document}

\maketitle

\begin{abstract}%

In this paper,  we present the demonstration of training a four-layer neural network entirely using fully homomorphic encryption (FHE), supporting both single-output and multi-output classification tasks in a non-interactive setting. A key contribution of our work is identifying that replacing \textit{Softmax} with \textit{Sigmoid}, in conjunction with the Binary Cross-Entropy (BCE) loss function, provides an effective and scalable solution for homomorphic classification. Moreover, we show that the BCE loss function, originally designed for multi-output tasks, naturally extends to the multi-class setting~\cite{kornblith2021better}, thereby enabling broader applicability. We also highlight the limitations of prior loss functions such as the SLE loss~\cite{chiang2023privacy3layer, chiang2023privacy} and the one proposed in the 2019 CVPR Workshop~\cite{nandakumar2019towards}, both of which suffer from vanishing gradients as network depth increases. To address the challenges posed by large-scale encrypted data, we further introduce an improved version of the previously proposed data encoding scheme, \textit{Double Volley Revolver}~\citep{chiang2022novel}, which achieves a better trade-off between computational and memory efficiency, making FHE-based neural network training more practical. The complete, runnable C++ code to implement our  work can be found at: \href{https://github.com/petitioner/ML.NNtraining}{$\texttt{https://github.com/petitioner/ML.NNtraining}$}. 

\end{abstract}

\listoftodos

\section{Introduction}

\subsection{Background}
Deep neural networks are a versatile and powerful tool with diverse applications, ranging from speech recognition to computer vision. The process of utilizing these networks typically involves two main phases: training and inference. In the training phase, an appropriate dataset is selected, and a network architecture is designed. The data is then used to learn the network weights, a process that may take days. Once the weights are stable and the network generates meaningful results, it can be used for inference, where it makes predictions on new, unseen data. While training can be time-consuming, inference is expected to be much faster.

However, many scenarios involve sensitive data that cannot be freely shared. For example, credit card transaction data is proprietary to credit card companies, and healthcare data, such as patient records, is confined to hospitals and unavailable to researchers studying conditions like cancer progression. Moreover, privacy regulations, such as the European GDPR, further restrict the availability of such data. Often, data owners lack the expertise to build deep learning models themselves, but privacy and confidentiality concerns prevent sharing the data with external providers.

Fully homomorphic encryption (FHE) offers a potential solution to balance the need for privacy with the utility of the data. Despite initial skepticism about its feasibility, FHE has seen significant advancements over the past decade, with various algorithmic improvements enabling research prototypes to demonstrate its application in different contexts. Although current implementations are still considered too slow for training deep neural networks, ongoing progress suggests that FHE could eventually be a viable approach for secure machine learning.

\subsection{Related Work}
While privacy-preserving machine learning has been explored for nearly two decades, relatively little research has focused on the specific application of homomorphic encryption within neural networks. To the best of our knowledge, the only notable prior work utilizing non-interactive homomorphic encryption in this context is CryptoNets by Gilad-Bachrach et al., which demonstrated a carefully constructed neural network capable of performing inference directly on encrypted data. This model achieved 99\% accuracy on the MNIST optical character recognition task and an amortized throughput of approximately 60,000 predictions per hour. More recently, fully homomorphic encryption has also been employed in the design of protocols for secure face matching and secure $k$-nearest neighbor search.

There has been significantly more progress in leveraging homomorphic encryption (HE) in combination with interactive secure computation protocols for neural networks. Early contributions in this direction include the works of Barni et al. and Orlandi et al., which employed additively homomorphic encryption together with interactive protocols, enabling inference on small networks within approximately ten seconds. More recently, a series of interactive protocols have been proposed for secure inference, such as SecureML by Mohassel and Zhang, MiniONN by Liu et al., Chameleon by Riazi et al., and GAZELLE by Juvekar et al.. The latter achieves inference on MNIST in as little as 30ms, and on CIFAR-10 in just 13 seconds.

However, all of these studies focus solely on the inference phase; none address the problem of privacy-preserving training. To the best of our knowledge, there is few prior work tackling the private training of neural networks. This absence is likely due to the prevailing belief that training under homomorphic encryption would be prohibitively slow. Additionally, training complex models often involves conditional operations such as comparisons and selections, which were traditionally deemed impractical using HE alone.

In this work, we take a further step toward challenging this perception by demonstrating that even non-interactive homomorphic encryption can support training in certain restricted scenarios. Previous research has explored privacy-preserving training and inference for simpler models, including linear regression and logistic regression.

\subsection{Contributions}

In this paper, we propose leveraging Fully Homomorphic Encryption (FHE) to facilitate the training of neural network models on encrypted data. This approach allows users to encrypt their data using a private (secret) key and share the encrypted data with a service provider. The service provider can then train the model without accessing the underlying data. Since the trained model is also encrypted, the service provider remains unaware of both the data and the learned model parameters. Furthermore, the resulting model is only useful to users with access to the private key, preventing unauthorized sharing. This method is particularly suited for scenarios where data owners wish to outsource deep learning tasks to external providers who possess the necessary expertise and computational resources, while ensuring that these providers do not gain undue advantages from the data or the model.

Our contributions are threefold:

\begin{enumerate}

\item We demonstrate that the Binary Cross-Entropy (BCE) loss function is effective for addressing single-output classification tasks in privacy-perserving machine learning. Given that BCE is inherently designed for multi-output scenarios, the proposed approach naturally extends to multi-output classification problems as well~\cite{kornblith2021better}. To the best of our knowledge, this is the first work to tackle multi-output classification in a non-interactive manner over data encrypted with fully homomorphic encryption (FHE).

\item We propose an improved version of the previously introduced data encoding scheme~\citep{chiang2022novel}, \emph{Double Volley Revolver}, which is better suited for large-scale neural network training. Our method achieves a favorable balance between the time and space complexity of the resulting algorithm.

\item Finally, we address the challenge of speeding up FHE computations through the smart implementation of ciphertext packing. Although ciphertext packing is a known technique in FHE, we utilize it strategically to minimize the number of bootstrapping operations and enable parallel computation across neurons, leading to significant reductions in computational complexity.

\end{enumerate}

\section{Preliminaries}

\subsection{Fully Homomorphic Encryption}

Homomorphic Encryption (HE) refers to a class of encryption schemes that support computation directly on encrypted data, without requiring access to the secret key. A scheme is called \emph{fully} homomorphic if it supports both addition and multiplication operations, thereby enabling arbitrary computations over ciphertexts. Since Gentry's groundbreaking work in 2009~\cite{gentry2009fully}, which introduced the first fully homomorphic encryption (FHE) scheme, there has been significant progress in the field. For instance, Brakerski, Gentry, and Vaikuntanathan~\cite{brakerski2014leveled} proposed the BGV scheme, a leveled FHE scheme that significantly improves efficiency. Smart and Vercauteren~\cite{SmartandVercauteren_SIMD} introduced a key optimization known as the \emph{Single Instruction Multiple Data} (SIMD) technique, which enables encrypting multiple plaintext slots into a single ciphertext via polynomial Chinese Remainder Theorem (CRT) packing.

Another major advancement, especially for machine learning applications, is the \emph{rescaling} technique~\cite{cheon2017homomorphic}, which allows control over the magnitude of plaintexts during computation, helping manage precision and ciphertext noise.

Modern FHE schemes such as \texttt{HEAAN} support a set of standard homomorphic operations, including:
\begin{itemize}
    \item $\texttt{Enc}$: Encryption of a plaintext vector;
    \item $\texttt{Dec}$: Decryption of a ciphertext;
    \item $\texttt{Add}$ and $\texttt{Mult}$: Homomorphic addition and multiplication of ciphertexts;
    \item $\texttt{cMult}$: Multiplication of a ciphertext with a constant plaintext vector;
    \item $\texttt{ReScale}$: Rescaling operation to reduce the plaintext scale;
    \item $\texttt{Rot}$: Rotation of encrypted vectors (e.g., circular shift);
    \item $\texttt{bootstrap}$: Bootstrapping to refresh ciphertexts and reduce accumulated noise.
\end{itemize}

These operations form the foundation for implementing non-trivial encrypted computations, including privacy-preserving machine learning tasks.

\subsubsection{Data Encoding}

To optimize homomorphic computations on databases, Kim et al.~\cite{IDASH2018Andrey} proposed an efficient encoding method that maximizes resource utilization in HE systems. Specifically, given a matrix-form database $Z$, the data is first linearized into a vector $V$ using row-wise flattening, which is then encrypted to produce the ciphertext $Z = \texttt{Enc}(V)$. 

Based on this vectorized encoding, two key operations are enabled by shifting the encrypted vector:
\begin{itemize}
    \item \textbf{Complete row shifting}: rotates entire rows within the matrix;
    \item \textbf{Incomplete column shifting}: simulates column-wise operations by partially rotating values across rows.
\end{itemize}

These techniques enable matrix manipulations directly in the encrypted domain, yielding modified matrices $Z'$ and $Z''$ respectively.

\begin{equation*}
 \begin{aligned}
 Z &= 
\left[ \begin{array}{cccc}
 x_{10}  &   x_{11}  &  \ldots  &  x_{1d}  \\
 x_{20}  &   x_{21}  &  \ldots  &  x_{2d}  \\
 \vdots         &   \vdots         &  \ddots  &  \vdots         \\
 x_{n0}  &  x_{n1}   &  \ldots  &  x_{nd}  \\
 \end{array}
 \right], 
& Z^{'}  = Enc
\left[ \begin{array}{cccc}
 x_{20}  &   x_{21}  &  \ldots  &  x_{2d}  \\
 \vdots         &   \vdots         &  \ddots  &  \vdots         \\
 x_{n0}  &   x_{n1}  &  \ldots  &  x_{nd}  \\
 x_{10}  &  x_{11}   &  \ldots  &  x_{1d}  \\
 \end{array}
 \right],   \\
 Z^{''}  &= Enc
\left[ \begin{array}{cccc}
 x_{11}  &  \ldots  &  x_{1d}  &   x_{20}  \\
 x_{21}  &  \ldots  &  x_{2d}  &   x_{30}  \\
 \vdots         &   \vdots         &  \ddots  &  \vdots         \\
 x_{n1}  &  \ldots  &  x_{nd}  &   x_{10} \\
 \end{array}
 \right], 
 & Z^{'''} = Enc
\left[ \begin{array}{cccc}
 x_{11}  &  \ldots  &  x_{1d}  &   x_{10}  \\
 x_{21}  &  \ldots  &  x_{2d}  &   x_{20}  \\
 \vdots         &   \vdots         &  \ddots  &  \vdots         \\
 x_{n1}  &  \ldots  &  x_{nd}  &   x_{n0} \\
 \end{array}
 \right]  .
 \end{aligned}
\end{equation*}

Moreover, complete column shifting to generate the matrix $Z^{'''}$ can be realized using two $\texttt{Rot}$ operations, two $\texttt{cMult}$ operations, and one $\texttt{Add}$ operation.

Subsequent works~\cite{han2018efficient, chiang2022novel} adopting the same encoding approach have introduced additional procedures such as $\texttt{SumRowVec}$ and $\texttt{SumColVec}$, which compute the summation of each row and column, respectively. These basic yet essential operations serve as foundational building blocks for more complex computations, including the homomorphic evaluation of gradients.

\subsection{Convolutional Neural Network}
Convolutional Neural Networks (CNNs) are a class of artificial neural networks inspired by biological visual systems. They are particularly well-suited for analyzing visual data and have demonstrated state-of-the-art performance in image recognition tasks. Notably, CNNs are among the few deep learning architectures that draw structural inspiration from the visual cortex in the human brain.

\subsubsection{Transfer Learning}
Transfer learning refers to techniques where a model pre-trained on one task is reused or adapted for another, typically related, task. In real-world scenarios, training entire CNNs from scratch is uncommon due to the limited availability of sufficiently large datasets. Instead, practitioners often employ a pre-trained CNN as a fixed feature extractor, transferring its learned representations to new tasks.

In our case, we freeze the weights of all layers in the selected pre-trained CNN except for the final fully connected layer. We then replace this final layer with a new one initialized with random weights (e.g., zeros), and train only this layer. This approach simplifies CNN training to that of multiclass logistic regression.

\paragraph{REGNET\_X\_400MF} 
For our privacy-preserving CNN training, we adopt a recent model design paradigm introduced by Facebook AI researchers, known as $\texttt{RegNet}$. This framework defines a low-dimensional design space consisting of simple and regular networks. We specifically select $\texttt{REGNET\_X\_400MF}$ as our fixed feature extractor and replace its final fully connected layer with a new one initialized with zero weights. The rest of the network remains frozen during training.

Since $\texttt{REGNET\_X\_400MF}$ is designed to accept color images of size $224 \times 224$, any grayscale input images are transformed by stacking them across three channels. Furthermore, input images of varying sizes are resized to the required dimensions. These preprocessing steps are performed using PyTorch utilities.

\subsection{Squared Likelihood-Error Loss}

Due to the inherent uncertainty and complexity of the Softmax function, directly approximating it in the encrypted domain is often impractical. To address this challenge, Chiang et al.~\cite{chiang2023privacy3layer, chiang2023privacy} adopt a classical mathematical strategy: transforming a difficult problem into a simpler one. Specifically, instead of approximating the Softmax function, they focus on the Sigmoid function, which has been extensively studied under encryption, particularly using least-squares approximation techniques.

To remain consistent with the typical use of Softmax in the log-likelihood loss function, they propose maximizing the following reformulated objective:
\[
L_1 = \prod_{i=1}^{n} \frac{1}{1 + \exp(-\mathbf{x}_i \cdot \mathbf{w}_{[y_i]}^\intercal)}.
\]
Empirical evaluations suggest that $\ln L_1$ performs suboptimally. A key reason is that its gradient and Hessian, for each individual example, involve only the class weight associated with that example, without accounting for other classes.

To overcome this limitation, Chiang et al.~\cite{chiang2023privacy3layer, chiang2023privacy}  introduce a new loss function:
\[
L_{11} = \prod_{i=1}^n \prod_{j=0}^{c-1} \left( 1 - \left(\bar{y}_{ij} - \text{Sigmoid}(\mathbf{x}_i \cdot \mathbf{w}_{[j]}^\intercal) \right) \right)^2.
\]
Initially, they attempt to maximize its logarithmic form:
\[
\ln L_{11} = \sum_{i=1}^n \sum_{j=0}^{c-1} \ln \left| 1 - \left(\bar{y}_{ij} - \text{Sigmoid}(\mathbf{x}_i \cdot \mathbf{w}_{[j]}^\intercal) \right) \right|.
\]
To better align with conventional learning paradigms, the formulation is further revised to minimize the logarithm of a new loss function $L_2$:
\[
L_2 = \prod_{i=1}^n \prod_{j=0}^{c-1} \left( \bar{y}_{ij} - \text{Sigmoid}(\mathbf{x}_i \cdot \mathbf{w}_{[j]}^\intercal) \right)^2,
\]
\[
\ln L_2 = \sum_{i=1}^n \sum_{j=0}^{c-1} \ln \left| \bar{y}_{ij} - \text{Sigmoid}(\mathbf{x}_i \cdot \mathbf{w}_{[j]}^\intercal) \right|.
\]

This new loss, referred to as the \textbf{Squared Likelihood Error (SLE)}, is empirically shown to perform competitively with the Softmax-based log-likelihood loss. Notably, the SLE formulation resembles the Mean Squared Error (MSE): whereas MSE sums squared errors, SLE takes the product of squared likelihood errors across all classes and examples.

A subsequent study~\cite{chiang2023privacy} extends this formulation to neural networks with one hidden layer. The proposed generalized loss is:
\[
L_3 = \sum_{i=1}^n \sum_{j=0}^{c-1} \left( \bar{y}_{ij} - \text{Sigmoid}(\mathbf{x}_i \cdot \mathbf{w}_{[j]}^\intercal) \right)^2.
\]
This reformulation offers greater interpretability by treating classification as a special case of regression. The use of Sigmoid activation in the output layer improves training stability and reduces the likelihood of divergence caused by inappropriate learning rates—a common issue in standard regression settings.

For first-order optimization methods such as gradient descent/ascent, it is standard practice to average the loss, yielding the \textbf{Mean Squared Likelihood Error (MSLE)}:
\[
L_2 = \frac{1}{n} \sum_{i=1}^n \sum_{j=0}^{c-1} \left( \bar{y}_{ij} - \text{Sigmoid}(\mathbf{x}_i \cdot \mathbf{w}_{[j]}^\intercal) \right)^2,
\]
\[
\ln L_2 = \frac{1}{n} \sum_{i=1}^n \sum_{j=0}^{c-1} \ln \left| \bar{y}_{ij} - \text{Sigmoid}(\mathbf{x}_i \cdot \mathbf{w}_{[j]}^\intercal) \right|.
\]

\section{Technical Details}
%

\subsection{Binary Cross-Entropy Loss}
Binary Cross Entropy is a widely-used loss function in machine learning, particularly in binary classification tasks. It measures the dissimilarity between the true labels and the predicted probabilities for each class, utilizing the concept of entropy from information theory. This loss function is especially effective when the output is a probability value indicating the likelihood of a particular class, and the model outputs a value in the range [0, 1] for each instance. Binary Cross-Entropy loss originates from the concept of cross entropy in information theory, first introduced by Shannon (1948), and is widely adopted in modern neural networks as the negative log-likelihood under a Bernoulli distribution.

The Binary Cross-Entropy Loss function is defined as:
\[
L = -\frac{1}{n} \sum_{i=1}^{n} \left[ y_i \log(p_i) + (1 - y_i) \log(1 - p_i) \right]
\]
where:
\begin{itemize}
    \item \( n \) is the total number of samples,
    \item \( y_i \) is the true label (0 or 1) of the \(i\)-th sample,
    \item \( p_i \) is the predicted probability for the positive class (class 1),
    \item \( \log \) represents the natural logarithm.
\end{itemize}

The Binary Cross-Entropy Loss quantifies the error between the predicted probability \( p_i \) and the true label \( y_i \). The logarithmic terms ensure that the penalty increases as the predicted probability diverges from the actual label. This loss is minimized during training, leading the model to improve its accuracy in classifying binary outcomes.

Had the formulation
\[
\ln L_{11} = \sum_{i=1}^n \sum_{j=0}^{c-1} \ln \left| 1 - \left(\bar{y}_{ij} - \text{Sigmoid}(\mathbf{x}_i \cdot \mathbf{w}_{[j]}^\intercal) \right) \right|
\]
been properly transformed into its dual form from the beginning, the resulting expression would naturally coincide with the Binary Cross-Entropy (BCE) loss:
\[
L_{\text{BCE}} = -\sum_{i=1}^n \sum_{j=0}^{c-1} \ln \left| 1 - \bar{y}_{ij} + \text{Sigmoid}(\mathbf{x}_i \cdot \mathbf{w}_{[j]}^\intercal)  \right|.
\]

Both the SLE loss function and the loss formulation adopted in the 2019 CVPR Workshop baseline~\cite{nandakumar2019towards} suffer from inherent limitations. In particular, they are prone to the vanishing gradient problem as the network depth increases, leading to ineffective training in more complex architectures. In contrast, the Binary Cross-Entropy (BCE) loss offers a more stable and scalable solution, making it a more suitable choice for deeper neural networks and ultimately serving as a more robust optimization objective.

\subsection{Double Volley Revolver}
Unlike those efficient, complex encoding methods~\citep{kim2018matrix}, $\texttt{Volley Revolver}$~\citep{chiang2022novel} is a simple, flexible matrix-encoding method specialized for privacy-preserving machine-learning applications, whose basic idea in a simple version is to encrypt the transpose of the second matrix for two matrices to perform multiplication.

The encoding method actually plays a significant role in implementing privacy-preserving CNN training. Just as Chiang mentioned in~\citep{chiang2022novel}, we show that Volley Revolver can indeed be used to implement homomorphic CNN training. This simple encoding method can help to control and manage the data flow through ciphertexts.

However, we don't need to stick to encrypting the transpose of the second matrix. Instead, either of the two matrices is transposed  would do the trick: we could also encrypt the transpose of the first matrix, and the corresponding multiplication algorithm due to this change is similar to the Algorithm $2$ from~\citep{chiang2022novel}.

Also, if each of the two matrices are too large to be encrypted into a single ciphertext, we could also encrypt the two matrices into two teams $A$ and $B$ of multiple ciphertexts. In this case, we can see this encoding method as $\texttt{Double Volley Revolver}$, which has two loops: the outside loop deals with the calculations between ciphertexts from two teams while the inside loop literally calculates two sub-matrices encrypted by two ciphertexts $A_{[i]}$ and $B_{[j]}$ using the raw algorithm of Volley Revolver.  

\subsubsection{Vertical Partitioning}

\begin{figure}[htp]
\centering
\includegraphics[scale=0.6]{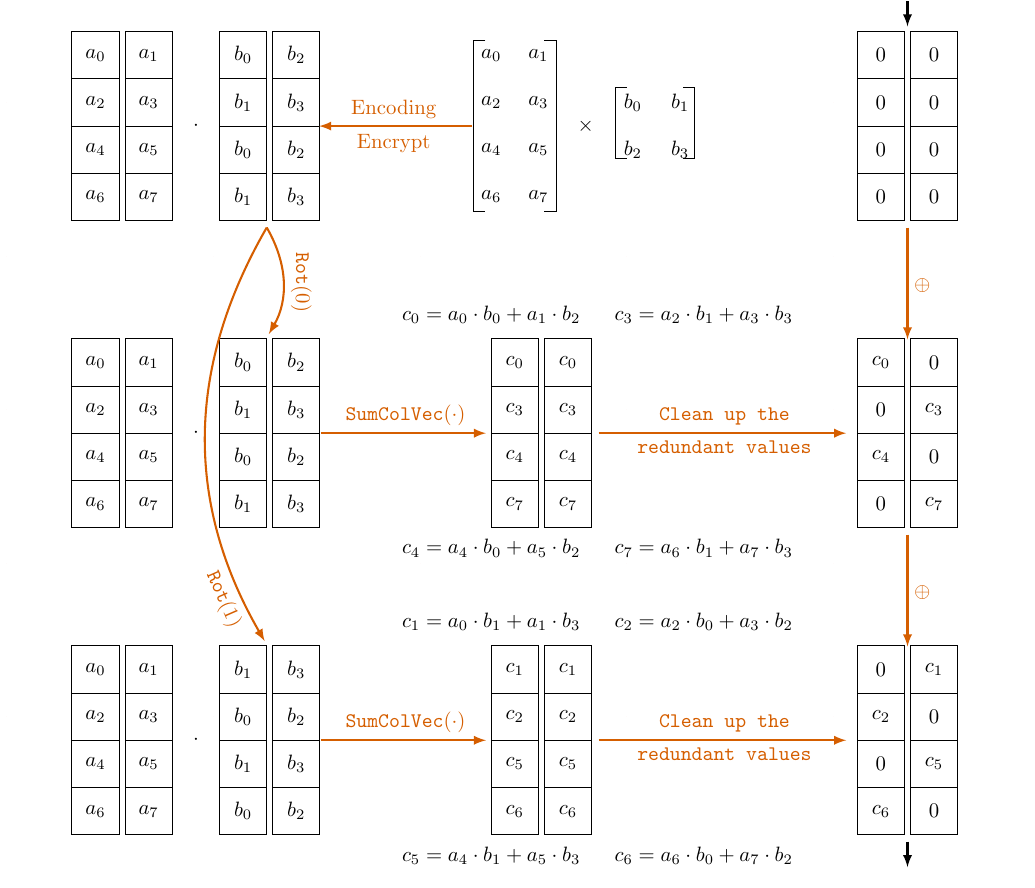}
\caption{
 The matrix multiplication algorithm of $\texttt{Volley Revolver}$ for the $4 \times 2$ matrix $A$ and the matrix $B$ of size $2 \times 2$ }
\label{ Matrix Multiplication }
\end{figure}

Figure~\ref{ Matrix Multiplication }  describes a simple case for the algorithm adopted in this encoding method.

\subsection{Deep Learning Model}

Deep learning models are commonly structured as multilayer neural networks, enabling the hierarchical extraction of abstract features through nonlinear transformations of lower-level representations, beginning with the raw input data. Figure~\ref{ 4LayerNN } illustrates a typical feedforward neural network architecture consisting of two hidden layers.

Each node (neuron) in the network computes its output by applying a nonlinear activation function to the weighted sum of its inputs, where a bias term---typically fixed at 1---is also included. Formally, the activation vector of the $\ell$-th layer, denoted $a^{(\ell)}$, is computed as:
\begin{equation}
    a^{(\ell)} = f\left(W^{(\ell)} a^{(\ell-1)}\right),
\end{equation}
where $f(\cdot)$ is the activation function, $W^{(\ell)}$ is the weight matrix for layer $\ell$, and $L$ denotes the total number of layers in the network.

Given a labeled training dataset $\{(x_i, y_i)\}_{i=1}^{N}$, the objective is to learn the set of weight matrices $\{W^{(\ell)}\}$ that minimizes a predefined loss function $\mathcal{L}$. This task corresponds to solving a non-convex optimization problem, which is typically addressed using variants of gradient descent.

In this work, we employ the widely-used \emph{Nesterov’s Accelerated Gradient} (NAG) algorithm. One full pass over the entire dataset is referred to as an \emph{epoch}. The gradient update is iteratively performed until convergence to a local minimum or until a maximum number of epochs is reached. The weight update rule is given by:
\begin{equation}
    W^{(\ell)} \leftarrow W^{(\ell)} - \alpha \frac{\partial \mathcal{L}_B}{\partial W^{(\ell)}},
\end{equation}
where $\mathcal{L}_B$ denotes the loss evaluated over the mini-batch $B$, and $\alpha$ is the learning rate.

During training, the forward pass computes the output of the network, and the loss at the output layer is determined accordingly. The backpropagation algorithm is then used to propagate the error backward through the network, allowing gradients to be computed for all layers.

\paragraph{ Nesterov’s Accelerated Gradient } With $\nabla$  or $\nabla^2$, first-order gradient algorithms or second-order Newton–Raphson method are commonly applied in MLE to maxmise $\ln L$. In particular, Nesterov’s Accelerated Gradient (NAG) is a practical solution for homomorphic MLR without frequent inversion operations.
It seems plausible that the NAG method is probably the best choice for privacy-preserving model training. 

\subsubsection{4-Layer Neural Networks}
In our implementation, we employ a 4-layer neural network consisting of two single hidden layers, following the same architecture as the baseline methods.

\begin{figure}[htp]
\centering
\includegraphics[scale=1.0]{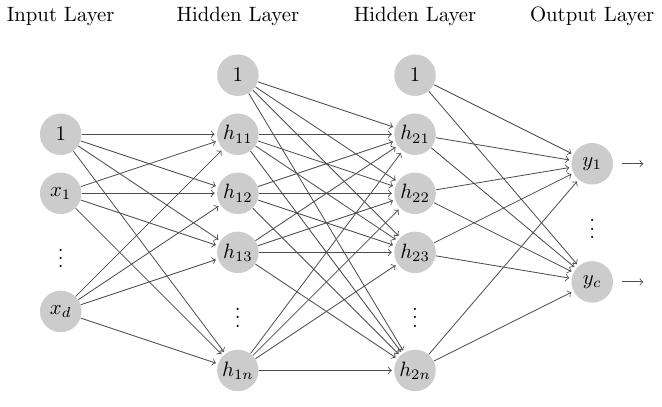}
\caption{
 A typical neural network with two hidden layers is illustrated, where black circles represent bias nodes that constantly emit a value of 1. The weight matrices \( W^{(\ell)} \) determine the contribution of each input signal to the activation function at each node. }
\label{ 4LayerNN }
\end{figure}

Figure~\ref{ 4LayerNN } illustrates a typical neural network comprising two hidden layers. The output of each node (or neuron) is computed by applying a non-linear activation function to a weighted sum of its inputs, which includes a constant bias term with value 1. The output vector of neurons in layer \( \ell \) (where \( \ell = 1, 2, \dots, L \)) is given by:
\[
\mathbf{a}^{(\ell)} = f\left( W^{(\ell)} \mathbf{a}^{(\ell-1)} \right),
\]
where \( f \) denotes the activation function, \( W^{(\ell)} \) is the weight matrix for layer \( \ell \), and \( L \) is the total number of layers in the network.

\begin{table}[htbp]
\centering
\caption{Result of machine learning on encrypted data}
\label{tab1}
\begin{tabular}{|l||l|c|}
\hline
Layer  & BaseLine~\cite{nandakumar2019towards} & Our \\
\hline
\hline
\multirow{1}{*}{Input} &   \multirow{1}{*}{nn.Reshape(64)} & \multirow{1}{*}{    } \\
\hline
FC-1 &  nn.Linear(64->32)       &  \\
\hline
ACT-1  & nn.Sigmoid             & quadratic/cubic polynomial activation \\
\hline
FC-2 & nn.Linear(32->16)        &  \\
\hline
ACT-2 & nn.Sigmoid              & low-degree polynomial activation function \\
\hline
FC-3 & nn.Linear(16->10)        &   \\
\hline
Output & nn.Sigmoid             &  polynomial sigmoid approximation \\
\hline
Sigmoid Approximation & homomorphic table lookup~\cite{IDASH2018gentry}            &  domain extension polynomials~\cite{cheon2022efficient} \\
\hline
Optimization Algorithm & Stochastic Gradient Descent           &  Nesterov’s Accelerated Gradient \\
\hline
\end{tabular}
\end{table}

Nandakumar et al.~\cite{nandakumar2019towards} actually use Mini-batch Gradient Descent instead of  Stochastic Gradient Descent.

\section{Homomorphic Training} Implementation/Homomorphic Implementation

\subsection{Polynomial Approximation}

Several established techniques exist for approximating nonlinear functions using polynomials. Classical methods such as \emph{Taylor expansion} and \emph{Lagrange interpolation} offer precise local approximations around a specific point. However, their accuracy deteriorates rapidly outside the vicinity of the expansion point, resulting in significant approximation error over broader intervals.

In contrast, the \emph{least squares approximation} method seeks to minimize the overall approximation error across a global domain, thereby providing more reliable performance over wider ranges. Due to its robustness and generality, it has been extensively adopted in practical applications, as demonstrated in. Both Python and MATLAB provide built-in functions---\texttt{polyfit(·)}---that implement least squares polynomial fitting for non-polynomial functions.

Another widely used technique is the \emph{minimax approximation}, which aims to minimize the maximum error over the approximation interval. This method ensures uniform approximation quality and is particularly suitable for scenarios where worst-case error bounds are critical.

Recent research has also focused on polynomial approximation over large intervals. For example, Cheon et al.~\cite{cheon2022efficient} proposed the use of \emph{domain extension polynomials}, which enable iterative extension of the approximation domain. This approach allows efficient approximation of sigmoid-like functions over significantly wider intervals and has proven effective for \emph{homomorphic evaluation}. In our work, we adopt their method to approximate the sigmoid function over the interval $[-64, 64]$.

\subsection{Homomorphic Evaluation }

\subsection{Computational Complexity} 
We conducted experiments for encrypted-domain processing on a dual-socket Intel Xeon E5-2698 v3 (Haswell architecture) server, featuring 16 cores per socket running at 2.30GHz and equipped with 250~GB of main memory. Compilation was performed using GCC~7.2.1, with NTL version~10.5.0 and GMP version~6.0 for arithmetic support.

\textbf{Future improvements:} These preliminary results demonstrate the feasibility of SGD training in the encrypted domain. Nevertheless, this work represents an initial effort, and further optimization opportunities remain. In particular, we have only begun to explore efficient batching/packing strategies. At present, we batch the inputs but still allocate a separate ciphertext for each weight parameter.

\section{Experiments}
The C++ source code to implement the experiments in this section  is openly available at: \href{https://github.com/petitioner/ML.NNtraining}{$\texttt{https://github.com/petitioner/ML.NNtraining}$} .

\subsection{Comparison with Baseline Work~\cite{nandakumar2019towards}}

\subsection{Multi-output Classification Performance}

\subsection{Transfer Learning Application}

\paragraph{Datasets}  

In our experiments, we use three widely-used datasets: USPS, MNIST, and CIFAR-10. Table \ref{tabdatasets} summarizes the key characteristics of these datasets.

\begin{table}[htbp]
\centering
\caption{Characteristics of the datasets used in our experiments}
\label{tabdatasets}
\begin{tabular}{|c|c|c|c|c|c|c|c|c|}
\hline
Dataset &   \mysplit{No. of Samples  \\ (training)}   &   \mysplit{No. of Samples  \\ (testing)}   &  No. of Features   & No. of Classes  \\
\hline
USPS &    7,291   &   2,007   &  16$\times$16   & 10  \\
\hline
MNIST &   60,000   &   10,000   &  28$\times$28   & 10  \\
\hline
CIFAR-10 &   50,000   &   10,000   &  3$\times$32$\times$32   & 10  \\
\hline
\end{tabular}
\end{table}

\paragraph{Parameters}

For the training data, we use the first 128 MNIST training images, and the entire test dataset is used for evaluation. Both the training and testing images have been pre-processed with the pre-trained model $\texttt{REGNET\_X\_400MF}$, resulting in a new dataset where each example has a size of 401.

The parameters of $\texttt{HEAAN}$ used in our experiments are as follows: $logN = 16$, $logQ = 990$, $logp = 45$, and $slots = 32768$, which ensure a security level of $\lambda = 128$. For further details on these parameters, refer to \cite{IDASH2018Andrey}. We did not use bootstrapping to refresh the weight ciphertexts, which limits our algorithm to 2 iterations. Each iteration takes approximately 11 minutes. The maximum runtime memory required is around 18 GB. 

The 128 MNIST training images are encrypted into 2 ciphertexts. The client, who owns the private data, uploads these two ciphertexts, two ciphertexts encrypting the one-hot labels $\bar{Y}$, one ciphertext encrypting $\bar{B}$, and one ciphertext encrypting the weight matrix $W$ to the cloud. The initial weight matrix $W_0$ is set to the zero matrix. After 2 iterations of training, the resulting MLR model achieves a precision of 21.49\% and a loss of -147206, which is consistent with the Python simulation results.


\paragraph{Performance}

\section{Conclusion}

In this work, we implemented privacy-persevering 4-layer NN training based on mere HE techniques by presenting a faster HE-friendly algorithm.

\bibliography{HE.CNNtraining}
\bibliographystyle{apalike}  

\end{document}